\documentclass[aps,prb,reprint,superscriptaddress,showpacs,
amsmath,citeautoscript,flushbottom,floatfix]{revtex4-1}

\usepackage{graphicx}

\newcommand{\vc}[1]{\boldsymbol{#1}}

\setcitestyle{numbers,square}

\begin{document}
\title{Magnetic anisotropy in the Kitaev model systems 
Na$_\mathbf{2}$IrO$_\mathbf{3}$ and RuCl$_\mathbf{3}$}

\author{Ji\v{r}\'{\i} Chaloupka}
\affiliation{Central European Institute of Technology,
Masaryk University, Kamenice 753/5, 62500 Brno, Czech Republic}
\affiliation{Department of Condensed Matter Physics, Faculty of Science,
Masaryk University, Kotl\'a\v{r}sk\'a 2, 61137 Brno, Czech Republic}

\author{Giniyat Khaliullin}
\affiliation{Max Planck Institute for Solid State Research,
Heisenbergstrasse 1, D-70569 Stuttgart, Germany}

\begin{abstract}
We study the ordered moment direction in the extended Kitaev-Heisenberg model
relevant to honeycomb lattice magnets with strong spin-orbit coupling. We
utilize numerical diagonalization and analyze the exact cluster groundstates
using a particular set of spin coherent states, obtaining thereby quantum
corrections to the magnetic anisotropy beyond conventional perturbative methods.
It is found that the quantum fluctuations strongly modify the moment direction
obtained at a classical level, and are thus crucial for a precise
quantification of the interactions. The results show that the moment direction
is a sensitive probe of the model parameters in real materials. Focusing on
the experimentally relevant zigzag phases of the model, we analyze the
currently available neutron and resonant x-ray diffraction data on
Na$_2$IrO$_3$ and RuCl$_3$, and discuss the parameter regimes plausible 
in these Kitaev-Heisenberg model systems.
\end{abstract}

\date{\today}

\maketitle


\section{Introduction}

Due to their intermediate spatial extension, $d$-electrons in transition metal
compounds comprise both the localized and itinerant features. This duality is
manifested in a rich variety of metal-insulator transitions
\cite{Mot74,Ima98}. Even deep in the Mott-insulating phase, the $d$-electrons
partially retain their kinetic energy, by making virtual hoppings to the
neighboring sites and forming the covalent bonds. The internal structure of
these bonds is dictated by the orbital shape of $d$-electrons as well as by
Pauli principle and Hund's interactions among spins. This results in an
intimate link between the nature of chemical bonds (``orbital order'') and
magnetism \cite{Goo63}, which can be cast in terms of phenomenological
Goodenough-Kanamori rules. 

The Kugel-Khomskii models \cite{Kug82} form a theoretical framework where the
``spin physics'' and ``orbital chemistry'' are treated on equal footing.
A special feature of these models is that the $d$-orbital is spatially
anisotropic and hence cannot satisfy all the bonds simultaneously. In high
symmetry crystals, this results in a picture of fluctuating orbitals
\cite{Kha00,Kha05}, where the frustration among different covalent bonds is
resolved by virtue of their quantum superposition, lifting the orbital
degeneracy without a static order.

It might seem that a relativistic spin-orbit coupling, which lifts the 
orbital degeneracy already on a single ion level \cite{Goo63,Kug82}, will 
readily eliminate the orbital frustration problem. This coupling does 
indeed greatly reduce the initially large spin-orbital Hilbert space of 
$d$-ions, leaving often just a twofold degenerate Kramers level with an 
effective (``pseudo'') spin one-half \cite{Abr70}. It turns out, however, 
that the pseudospins still well ``remember'' the orbital frustration, by 
inheriting the bond-directional nature of orbital interactions via 
the spin-orbit entanglement \cite{Kha05}. 

The bond-directional nature of pseudospin interactions has profound
consequences for magnetism (as well as for the properties of doped systems
\cite{Kha04}). The most remarkable example, pointed out in
Ref.~\onlinecite{Jac09}, is a possible realization of the Kitaev's honeycomb
model \cite{Kit06} in materials with the $d^5(t_{2g})$ electronic
configuration such as Na$_2$IrO$_3$. This theoretical proposal has sparked a
broad interest in honeycomb lattice pseudospin systems (see the recent review
article \cite{Rau16} and references therein). 

There is a direct experimental evidence \cite{Chu15} that the Kitaev-type
interactions are indeed dominant in Na$_2$IrO$_3$. Unusual features pointing
towards the Kitaev model has been observed \cite{Ban16} also in spin
excitation spectra of RuCl$_3$ (this compound was suggested \cite{Plu14} to
host pseudospin physics, too). On the other hand, it is also clear that there
are terms in the pseudospin Hamiltonian that take these systems away from the
Kitaev spin-liquid phase window \cite{Cha10}. The identification of these
``undesired'' interactions and clarification of their dependence on material
parameters is an important issue that has been in the focus of many recent
studies.

Experimentally, the strength of a dominant Kitaev coupling $|K|$ can readily
be evaluated from an overall bandwidth of spin excitations; however, the
determination of its sign and quantification of the subdominant terms is not
straightforward and needs a theory support. The aim of this paper is to show
that the direction of the ordered moments, which can be extracted from the
neutron and \mbox{x-ray} diffraction data, contains a valuable information on 
the model parameters, including the sign of $K$. Considering a symmetry dictated
form of the model Hamiltonian, we calculate the pseudospin direction
fully including quantum fluctuations which are expected to be crucial in
frustrated spin models. We will point out that the pseudospin itself is not
directly probed by neutrons; rather, they detect the direction of the 
{\it magnetic} moment which is not the same as that of the pseudospin. 
Similarly, we will describe how to extract the pseudospin direction from 
resonant \mbox{x-ray} scattering (RXS) data. 

The paper is organized as follows. Section \ref{sec:model} introduces the
model Hamiltonian. Section \ref{sec:class} briefly discusses the pseudospin
easy axis direction on a classical level. Section \ref{sec:method} introduces
the method of deriving the moment direction from exact diagonalization (ED)
data. Section \ref{sec:EDresults} presents the ED results on moment direction
as a function of model parameters. Section \ref{sec:exper} considers a
relation between the pseudospins and magnetic moments probed by neutron
diffraction and RXS experiments, and discusses implications of the theory for
Na$_2$IrO$_3$ and RuCl$_3$. Appendix \ref{app:numer} compares the method of
Sec.~\ref{sec:method} with the standard approach. Appendix \ref{app:RXS}
derives the equations used in the analysis of RXS data. Finally, Appendix
\ref{app:trans} discusses how the trigonal field can be extracted from $J=3/2$
magnetic excitation spectra.



\section{Extended Kitaev-Heisenberg model}
\label{sec:model}

\begin{figure}
\begin{center}
\includegraphics[scale=1.00]{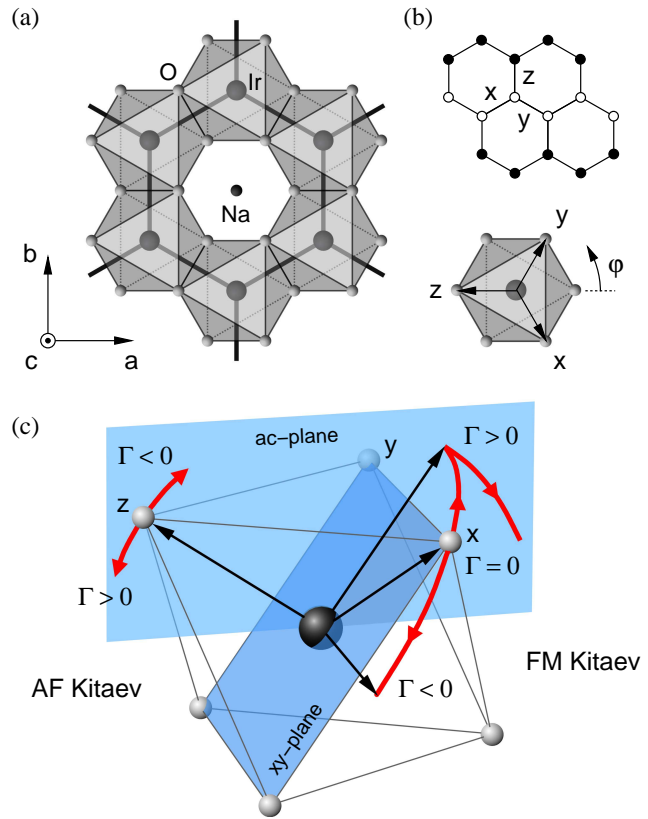}
\caption{(Color online)
(a)~Top view of the honeycomb lattice of the edge-shared IrO$_6$ octahedra in
\mbox{Na$_2$IrO$_3$}. 
(b) Three types of bonds and zigzag-AF state where \mbox{$x$-} and
\mbox{$y$-bonds} bonds connecting similar dots are FM, while the
\mbox{$z$-bonds} are AF (top), and the orientation of the cubic axes $x$, $y$,
$z$ with respect to the octahedra (bottom).
(c)~The possible directions of the ordered moment in the above zigzag state.
In the AF Kitaev case the moment is tied to the cubic \mbox{$z$-axis} and
deviates from it only slightly with nonzero $\Gamma$. In the FM Kitaev case
with $\Gamma=0$, it is constrained to the \mbox{$xy$-plane} classically, and
pinned to a cubic $x$- or \mbox{$y$-axis} when quantum fluctuations are
included.  Nonzero $\Gamma<0$ gradually pushes the moment direction towards
the \mbox{$b$-axis} in the honeycomb plane, while positive $\Gamma$ drives it
first towards the \mbox{$ac$-plane} [which is reached at $\Gamma\approx
0.05|K|$, see Fig.~\ref{fig:Gdep}(a)], and then rotates the moment within the
\mbox{$ac$-plane} towards the \mbox{$a$-axis}.
}
\label{fig:schem}
\end{center}
\end{figure}

To describe the interactions among the pseudospins (referred to as ``spins'' 
below), we adopt a model containing all symmetry allowed nearest-neighbor
(NN) terms and the longer-range Heisenberg interactions
\begin{equation}\label{eq:Hfull}
\mathcal{H} = 
\sum_{\langle ij \rangle \in \mathrm{NN}}
\mathcal{H}_{ij}^{(\gamma)}
+\!\!\!\sum_{\langle ij \rangle \notin\mathrm{NN}} 
\!\!\!J_{ij}\vc S_i\cdot \vc S_j \;.
\end{equation}
The nearest neighbor contribution is the extended Kitaev-Heisenberg model
\cite{Kat14,Rau14a,Rau14b} that, apart from the Heisenberg interaction,
includes all the bond-anisotropic interactions compatible with the symmetries
of a trigonally-distorted honeycomb lattice. Its $z$-bond contribution (see
Fig.~\ref{fig:schem} for the definitions of the bonds and spin axes) takes the
following form:
\begin{multline}\label{eq:Hcubicz}
\mathcal{H}_{ij}^{(z)} = K\, S_i^z S_j^z + 
J\, {\vc S}_i \cdot {\vc S}_j \\ 
\!+\!\Gamma(S_i^x S_j^y \!+\! S_i^y S_j^x)
\!+\!\Gamma'(S_i^x S_j^z \!+\! S_i^z S_j^x \!+\! S_i^y S_j^z \!+\! S_i^z S_j^y).
\end{multline}
The Hamiltonian contributions for the other bonds ($x$ and $y$) are obtained 
by a cyclic permutation among $S_x, S_y, S_z$. The resulting alternation of
the local easy axis directions from bond-to-bond, imposed by the Ising-like 
term $K$, brings about a strong frustration which, as discussed above, can 
be traced back to the orbital frustration problem in Kugel-Khomskii type 
models. An extensive discussion of the above Hamiltonian and its nontrivial 
symmetry properties can be found in Ref.~\onlinecite{Cha15}. 

With the Kitaev-coupling $K$ alone, the model has a spin-liquid ground state.
Both \mbox{Na$_2$IrO$_3$} and \mbox{RuCl$_3$} show spin order where the
zigzag-type ferromagnetic (FM) chains, running along $a$-direction, are
coupled to each other antiferromagnetically (AF), see Fig.~\ref{fig:schem}(b).
This order becomes a ground state of the Kitaev model with $K>0$ (AF sign),
when a small FM $J<0$ Heisenberg coupling is added \cite{Cha13}. If the Kitaev
coupling is negative, $K<0$ (FM sign), then zigzag order emerges due to
longer-range AF couplings \cite{Kim11,Cho12} and/or $\Gamma, \Gamma'$ terms
\cite{Rau14a,Rau14b,Cha15}. Given that the stability of the Kitaev-liquid
phase against perturbations strongly depends on the sign of $K$ \cite{Cha13},
which scenario is realized in a given compound becomes an important issue.

Leaving aside the ``orbital chemistry'' aspects that decide the sign of $K$ as
well as the other model parameters, we just mention that various ab-initio
estimates (see, e.g., \cite{Kat14,Yam14,Win16}) generally support FM $K<0$
regime, most likely reflecting the decisive role of Hund's coupling effect on
$K$ emphasized earlier \cite{Jac09,Cha10}. However, we take here a
phenomenological approach, considering the model with free parameter values
including both signs of $K$. The $J$, $\Gamma$, and $\Gamma'$ values are
varied such that the ground state stays within the zigzag phase. Based on
a recent result \cite{Win16} that third-NN Heisenberg coupling $J_3$ is more
significant than second-NN $J_2$ in both \mbox{Na$_2$IrO$_3$} and
\mbox{RuCl$_3$}, we replace $J_{ij}$ in \eqref{eq:Hfull} by $J_3$, reducing
thereby the parameter space.

The magnetic anisotropy in the present model is a nontrivial problem, since
the leading term $K$ is anisotropic by itself, and, on top of this highly
frustrated interaction, the other terms which eventually drive a magnetic
order in real compounds have a strong impact on magnetic energy profile. As
illustrated in Fig.~\ref{fig:schem}(c) and discussed in detail below, the
ordered moment direction is very sensitive to the model parameters, and it
shows a qualitatively different behavior in case of FM and AF Kitaev
couplings. We note that the ``moment direction'' in this figure refers to that
of pseudospin; Section VI explains how it is related to the magnetic moments
probed by neutron and \mbox{x-ray} diffraction experiments.



\section{Classical Moment direction}
\label{sec:class}

Let us briefly mention the results of a classical analysis (for details see
Appendix B of Ref.~\onlinecite{Cha15}) assuming the zigzag order with
antiferromagnetic $z$-bonds as shown in Fig.~\ref{fig:schem}(b).
On this level, the moment direction is determined solely by the anisotropy
parameters $K$, $\Gamma$, and $\Gamma'$ and corresponds to the eigenvector 
of the matrix
\begin{equation}
M=	
\begin{pmatrix}
2K               & -\Gamma+2\Gamma' & \Gamma \\
-\Gamma+2\Gamma' & 2K               & \Gamma \\
\Gamma           & \Gamma           & 0        
\end{pmatrix}
\end{equation}
that has the lowest eigenvalue. This minimizes the anisotropic contribution in
the classical energy per site of the zigzag phase,
$E_\mathrm{class}=\frac18(J-K-3J_3)+\frac18 \vc m^T M \vc m$,
where $\vc m$ is a unit vector. The dominant Kitaev interaction contributing
by the diagonal terms makes the main choice -- it prefers either the
\mbox{$xy$-plane} (FM $K<0$) or the \mbox{$z$-axis} (AF $K>0$). The smaller
$\Gamma$ and $\Gamma'$ terms lead to a finer selection of the ordered moment
direction.

In the case of the zigzag order stabilized by AF $K>0$ and FM $J<0$, the
ordered moment direction is close to the \mbox{$z$-axis} being slightly tilted
in the \mbox{$ac$-plane} mainly by virtue of $\Gamma$ [see
Fig.~\ref{fig:schem}(c)].

The FM $K<0$ case, where the zigzag order is stabilized by $\Gamma$ and $J_3$ 
terms, is more complex. With $\Gamma=\Gamma'=0$, the entire \mbox{$xy$-plane}
is degenerate on a classical level. Further selection depends on the sign of 
$\Gamma-2\Gamma'$, with the positive and negative sign making the moment to 
jump into the \mbox{$ac$-plane} or the \mbox{$b$-axis} in the honeycomb plane,
respectively. In the former case, an increasing $\Gamma$ further pushes 
the moment closer to the honeycomb plane. As it has been found earlier
\cite{Cha10,Siz16} and discussed below, the Kitaev term generates an 
additional magnetic anisotropy due to quantum and/or thermal fluctuations, 
pinning the moment direction to the cubic axes. This will turn the above 
jumps into a gradual rotation of the easy axis with changing $\Gamma$, 
along the path shown in Fig.~\ref{fig:schem}(c).



\section{Extraction of the moment direction from a cluster groundstate}
\label{sec:method}

To determine the groundstate of the Hamiltonian \eqref{eq:Hfull} and obtain
the moment direction as a function of model parameters more rigorously than in 
the previous perturbative methods, we have performed an exact diagonalization 
using a hexagon-shaped 24-site supercell covering the honeycomb lattice. This 
cluster is highly symmetric and compatible with all the hidden symmetries of 
the model \cite{Cha15} so that no bias induced by the cluster geometry 
is expected. 

Since the cluster groundstate does not spontaneously break the symmetry 
and corresponds to a superposition of all possible degenerate orderings, 
the identification of the ordered moment direction is not straightforward. 
One possibility is to evaluate the $3\times 3$ correlation matrix 
$\langle S^\alpha_{-\vc Q} S^\beta_{\vc Q}\rangle$ ($\alpha,\beta=x,y,z$)
at the ordering vector $\vc Q$ and to take the direction of the eigenvector
corresponding to its largest eigenvalue. Because of specific problems of this
standard approach in the present context (see Appendix \ref{app:numer} for 
details), we have developed here another method that brings a more intuitive 
picture of the exact groundstate by ``measuring'' the presence of the 
classical states with a varying moment direction. As a basic building block,
we utilize the spin-$\frac12$ coherent state
\begin{equation}
|\theta,\phi\rangle 
= \mathcal{R}_z(\phi) \mathcal{R}_y(\theta) |\!\uparrow\,\rangle
= \mathrm{e}^{-i \phi S^z} \mathrm{e}^{-i \theta S^y} |\!\uparrow\,\rangle
\end{equation}
that is fully polarized along $(\theta,\phi)$-direction \cite{Aue94}. Here
the cubic axes are used as a convenient reference frame and $\theta$, $\phi$
are the conventional spherical angles. A spin coherent state on the cluster
is constructed as a direct product
\begin{equation}
|\Psi\rangle = \prod_{j=1}^N |\theta_j,\phi_j\rangle
\end{equation}
with the unit vectors 
$\vc m_j=(\cos\phi\sin\theta,\sin\phi\sin\theta,\cos\theta)_j$
forming the desired pattern. In this fully polarized, classical state 
$\langle \Psi | S_i^\alpha S_j^\beta | \Psi \rangle = 
\frac14 m_i^\alpha m_j^\beta$ and the energy $\langle \Psi |
\mathcal{H} | \Psi \rangle$ is thus equal to the classical energy.
We consider only collinear states of FM, AF, and zigzag type. For
example, a FM state with the moment direction $(\theta,\phi)$ 
is explicitly expressed as
\begin{equation}\label{eq:cohFM}
|\Psi\rangle = \prod_{j=1}^N \left(
 \mathrm{e}^{-i\phi/2}\cos\tfrac\theta2 \, |\!\uparrow\,\rangle_j
+\mathrm{e}^{+i\phi/2}\sin\tfrac\theta2 \, |\!\downarrow\,\rangle_j
\right) \;.
\end{equation}
By varying $\theta$ and $\phi$ and evaluating the overlap with the exact
cluster groundstate $|\mathrm{GS}\rangle$, we obtain the probability map 
$P(\theta,\phi)=|\langle \Psi |\mathrm{GS}\rangle|^2$. The ordered moment 
direction is then identified by locating the maxima of $P(\theta,\phi)$.

There is an intrinsic width of the peaks in $P(\theta,\phi)$ due to the 
nonzero overlap of the spin coherent states, namely
$|\langle \Psi | \Psi' \rangle |^2=\cos^{2N}(\frac12\Omega)$, where 
$\Omega$ is the angle between the directions $(\theta,\phi)$ and
$(\theta',\phi')$. This gives an approximate half-width at half-maximum of 
$\sqrt{2/N}$ (in terms of the angular distance from the maximum), evaluating
to about $17^\circ$ for $N=24$. Despite this sizable intrinsic width, the
ordered moment direction can be detected with a high accuracy (limited only by
the accuracy of the groundstate vector), as we see below.



\section{Moment direction -- exact diagonalization results}
\label{sec:EDresults}

\subsection{Testing the method: nearly Heisenberg limit}
\label{sec:obd}

\begin{figure}
\begin{center}
\includegraphics[scale=1.00]{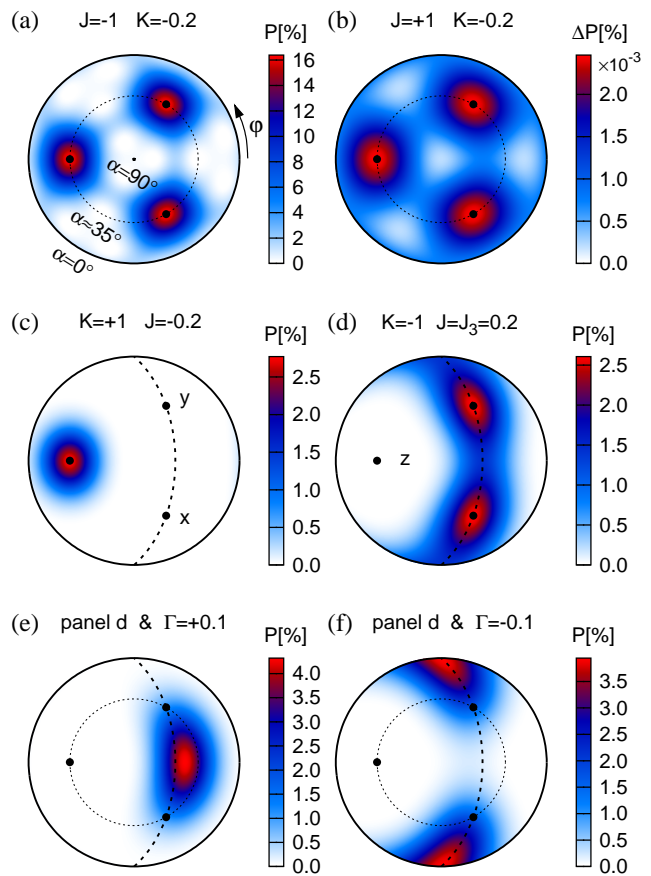}
\caption{(Color online)
(a)~Map of the probability of the spin coherent state given by
Eq.~\eqref{eq:cohFM} in the FM groundstate of the KH model near the Heisenberg
limit. The radial coordinate gives the angle $\alpha$ to the honeycomb plane,
the polar angle $\varphi$ matches that defined in Fig.~\ref{fig:schem}(b).
(b)~Probability map for the AF groundstate obtained using small $K$ and
dominant $J>0$. Only the variation $\Delta P$ on top of $P_0=2.923\%$ is
shown.
(c)~Probability map for the zigzag phase of the KH model with $K>0$, $J<0$
reveals a strong pinning to the \mbox{$z$-axis}. The coherent state
corresponding to the zigzag pattern in Fig.~\ref{fig:schem}(b) was used.
Directions lying in the \mbox{$xy$-plane} are indicated by the dashed line.
(d)~Soft \mbox{$xy$-plane} for FM $K<0$ zigzag stabilized by $J_3$. Cubic axes
$x$ and $y$ are selected but the moment strongly fluctuates in the plane.
(e,f)~The same as in panel (d) but extended by a sizable $\Gamma$-term forcing
the moment into the \mbox{$ac$-plane} (left) or the \mbox{$b$-axis} (right).
}
\label{fig:prob}
\end{center}
\end{figure}

Before discussing in detail the ordered moment direction in the zigzag phases,
relevant for actual compounds \mbox{Na$_2$IrO$_3$} and \mbox{RuCl$_3$}, let 
us demonstrate the above method by considering the Kitaev-Heisenberg model 
close to the Heisenberg limit, $|J|\gg|K|$, with both signs of $J$. In such a
situation, the FM or AF order is established by the dominant isotropic
interaction, while the anisotropic Kitaev interaction merely selects the 
easy axis direction via an order-from-disorder mechanism \cite{Tsv95}. 

We start with the FM case $J<0$. Presented Fig.~\ref{fig:prob}(a) is the 
corresponding probability map obtained by the method of previous 
Sec.~\ref{sec:method} for $K/J=0.2$. The probability is clearly peaked at 
the directions of the cubic axes attaining there the maximum value 
$P_\mathrm{max}$ slightly less than $\frac16$. This is due to the cluster 
groundstate being a superposition of six possible classical states and a small 
contribution of quantum fluctuations. The width of the peaks matches well
the intrinsic width estimated in Sec.~\ref{sec:method}.

That the $K$-term favors cubic axes for the ordered moment follows also 
from simple analytical calculations. By treating the quantum fluctuations 
within second order perturbation expansion (see Ref.~\onlinecite{Jac15} 
for details), we obtain the magnetic anisotropy energy 
\begin{equation}\label{eq:E2FM}
\delta E^{(2)}_{\mathrm{FM}} \approx
\frac{K^2}{64|J|} \left(1-m_x^4-m_y^4-m_z^4\right), 
\end{equation}
depending on the moment direction given by a unit vector 
$\vc m=(m_x,m_y,m_z)$. This quantum correction on top of the isotropic
classical energy is minimized for $\vc m$ pointing along the cubic axes
$x,y,z$ that become the easy axes, consistent with the ED result. 

The case of the AF $J>0$ is rather different due to the presence of large
quantum fluctuations already in the Heisenberg limit. This is manifested in an
almost flat probability profile with $P$ of about $3\%$ [see
Fig.~\ref{fig:prob}(b)]. Nevertheless, the probability maxima again precisely
locate the $x,y,z$ directions for the ordered moments, consistent with the
``order-from-disorder'' calculations \cite{Kha01,Cha10,Nus15,Jac15,Siz16} in
the models containing compass- or Kitaev-type bond-directional anisotropy.


\subsection{Moment direction in the zigzag phases}
\label{sec:zz}

Having verified the method, we now move to the zigzag phases observed in
Na$_2$IrO$_3$ and RuCl$_3$. We first inspect the case of $\Gamma, \Gamma'=0$
when the anisotropy is due to the Kitaev-term alone. Shown in
Fig.~\ref{fig:prob}(c) is the probability map for AF $K>0$ and FM $J<0$, where
the \mbox{$z$-axis} is selected already on the classical level as discussed in
Sec.~\ref{sec:class} \cite{noteZZ}. The probability is indeed strongly peaked at the
direction of the \mbox{$z$-axis}. The small $P_\mathrm{max}$ of about $3\%$
is again a signature of large quantum fluctuations in the groundstate. Note
that this number contains an overall reduction factor of $\frac16$ due
to the six possible zigzag states being superposed in the cluster groundstate.

The probability map Fig.~\ref{fig:prob}(d) for the FM $K<0$ zigzag case
reveals the moment being constrained to the vicinity of the \mbox{$xy$-plane},
as expected from classical considerations. Within this plane, the
order-from-disorder mechanism selects the cubic axes $x$ and $y$ where the
probability reaches its maxima. Concluding the survey of the probability
maps, we show $P$ calculated including a large enough $\Gamma$ that leads to
the selection of a direction within the \mbox{$ac$-plane} [$\Gamma>0$,
Fig.~\ref{fig:prob}(e)] or the \mbox{$b$-axis} [$\Gamma<0$,
Fig.~\ref{fig:prob}(f)].

\begin{figure}
\begin{center}
\includegraphics[scale=1.00]{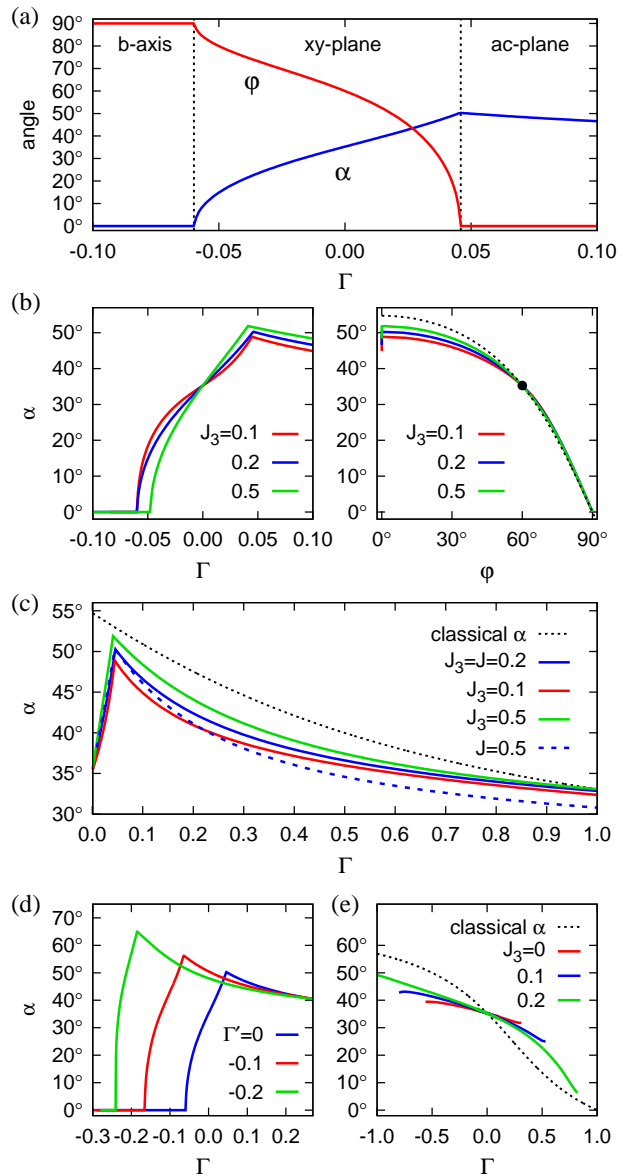}
\caption{(Color online)
(a)~$\Gamma$-dependent angles $\alpha$, $\varphi$ specifying the moment
direction reveal three regimes for FM $K$ zigzag supported by small $J_3$.
The values $K=-1$ and $J=J_3=0.2$ were used. At $\Gamma=0$, the angles give
the direction towards an oxygen ion. A crossover in the interval $|\Gamma|
\lesssim 0.05$ corresponds to the path shown in Fig.~\ref{fig:schem}(c).
(b)~Left panel shows the angle $\alpha$ for $K=-1$, $J=0.2$ and several $J_3$
values manifesting a stronger pinning to the cubic axis at smaller $J_3$. The
same data are presented as $\alpha(\varphi)$ in the right panel together with
$\alpha(\varphi)$ corresponding to the \mbox{$xy$-plane} (dashed). The black
dot indicates the cubic axis direction.
(c)~The angle $\alpha$ for larger values of $\Gamma>0$ compared to the
classical result of Ref.~\onlinecite{Cha15} (dotted). The blue solid curve is
a continuation of that of panel (a), red and green curves are calculated using
different $J_3$ values used in panel (b), blue dashed one for a larger $J$
value.
(d)~The angle $\alpha$ for the parameters $K=-1$, $J=J_3=0.2$ and several
$\Gamma'$ values.
(e)~$\Gamma$-dependent $\alpha$ in the AF $K=+1$ case with $J=-0.2$ and
several $J_3$ values compared to the classical result of
Ref.~\onlinecite{Cha15} (dotted). The endpoints of the curves are determined
by a sharp drop of the probability of the classical zigzag state indicating a
phase boundary.
}
\label{fig:Gdep}
\end{center}
\end{figure}

The above three examples for the FM $K$ zigzag indicate a rather complex 
behavior of
the moments in this case, as already suggested in Fig.~\ref{fig:schem}(c). In
the following, we therefore focus on the full $\Gamma$-dependence presented in
Fig.~\ref{fig:Gdep}(a) in the form of the angles $\alpha(\Gamma)$ (the angle
to the honeycomb plane) and $\varphi(\Gamma)$ (polar angle of the projection
into the honeycomb plane).
Instead of the jump in $\alpha(\Gamma)$ obtained on a classical level, we find
a finite window $|\Gamma|\lesssim 0.05|K|$ of an order-from-disorder
stabilized phase, where the moment direction gradually moves from the cubic
axis ($\Gamma=0$) to either \mbox{$b$-axis} ($\Gamma<0$) or to the \mbox{$ac$-plane}
($\Gamma>0$). Once the critical value of $\Gamma$ is reached, the moment
either stays along the \mbox{$b$-axis} or is pushed down within the 
\mbox{$ac$-plane} closer to the honeycomb plane.
Fig.~\ref{fig:Gdep}(b) illustrates the evolution of $\alpha(\Gamma)$ for
different values of $J_3$ stabilizing the zigzag order. For small $J_3$, the
dominant directional Kitaev term makes the moment more pinned to the cubic
axes, which is manifested by a significantly reduced slope of $\alpha(\Gamma)$
near $\Gamma=0$ compared to the large-$J_3$ case. On the other hand, the
critical values of $\Gamma$ are only slightly affected by $J_3$.

The above crossover behavior near $\Gamma=0$ may be easily understood and even
semi-quantitatively reproduced by considering a competition of the classical
energy and the order-from-disorder potential as follows.
Keeping the moment $\vc m=(\cos\phi,\sin\phi,0)$ within the \mbox{$xy$-plane}
preferred by $K<0$, we can evaluate the classical energy per site
\begin{equation}\label{eq:Eclxypl}
E_\mathrm{class} = \tfrac18(K-3J_3+J) 
- \tfrac18 (\Gamma-2\Gamma') \sin 2\phi \;.
\end{equation}
In this contribution, the anisotropy is due to the $\Gamma$- and
$\Gamma'$-terms only. $E_\mathrm{class}$ is complemented by an
order-from-disorder potential $E_\mathrm{fluct}(\phi)$ that should contain four
equivalent minima at $\phi=0,\frac12\pi,\pi,\frac32\pi$ corresponding to the
cubic axes (supported by the $K$ term). Such a potential can be represented
by the following form: 
\begin{equation}\label{eq:Efluct}
E_\mathrm{fluct}=V \sin^2 2\phi \;,
\end{equation}
approximating $E_\mathrm{fluct}(\phi)$ by its lowest harmonic.
This function is characterized by a single unknown parameter -- the barrier 
height $V$, determined mainly by the dominant $K$. Assuming $\Gamma'=0$, 
the minimization of the total energy $E_\mathrm{class}+E_\mathrm{fluct}$ 
gives $\phi(\Gamma)=\frac12\arcsin\frac{\Gamma}{16V}$
and the critical value $\Gamma_\mathrm{crit}=16V$. 
This enables us to extract effective $V$ from our numerical 
data. By taking $\Gamma_\mathrm{crit}\approx 0.05|K|$ observed in 
Fig.~\ref{fig:Gdep}(a,b) we get $V\approx 0.003|K|$.
Furthermore, converting $\phi$ in the \mbox{$xy$-plane} to the angle 
$\alpha$ to the honeycomb plane, we obtain ``phenomenological''
$\alpha(\Gamma)=\arcsin\sqrt{\frac13(1+\frac{\Gamma}{16V})}$
that roughly approximates the numerical $\alpha(\Gamma)$ data.
The agreement between these two $\alpha(\Gamma)$ profiles improves with
increasing $J_3$, when the order-from-disorder potential becomes more harmonic
and the deviation of the moment direction from the \mbox{$xy$-plane}
for $\Gamma>0$ reduces [see Fig.~\ref{fig:Gdep}(b)]. In fact, the above 
equations \eqref{eq:Eclxypl} and \eqref{eq:Efluct}, together with the value 
of $V\approx 0.003|K|$ extracted from the ED data, may be used for a 
semi-quantitative determination of the easy axis direction within the 
\mbox{$xy$-plane}.

For curiosity, we have evaluated the potential barrier $V$ also analytically,
by two slightly different methods. First, as in Sec.~\ref{sec:obd}, we
estimated quantum corrections for zigzag phase along the lines of
Ref.~\onlinecite{Jac15}. This reproduced the above form \eqref{eq:Efluct} of the
anisotropy potential, and provided a consistent estimate of $V\approx
0.005|K|$. An alternative evaluation of the anisotropy potential within the
linear spin-wave framework resulted in zero-point energy of the same form as
\eqref{eq:Efluct} again, but with an overestimated value of 
$V\approx 0.014|K|$.

In Na$_2$IrO$_3$ the moment direction was found \cite{Chu15} 
in the \mbox{$ac$-plane} suggesting
that $\Gamma>\Gamma_\mathrm{crit}$ for this material. We thus focus on this
particular case and investigate how the precise value of $\alpha$ is affected
by the model parameters in more detail. Already on a classical level, finite
$\Gamma>0$ rotates the moment within the \mbox{$ac$-plane} from 
$\alpha\approx 54.7^\circ$ (corresponding to the \mbox{$xy$-plane}) toward the 
honeycomb plane ($\alpha=0$). Such an effect is well visible also in 
Fig.~\ref{fig:Gdep}(a,b). Presented in Fig.~\ref{fig:Gdep}(c) are a few 
representative $\alpha(\Gamma)$ curves for larger values of $\Gamma$ up to 
$|K|$ that serve as a test of the classical prediction
\begin{equation}
\tan 2\alpha = 4\sqrt{2}\, \frac{1+r}{7r-2} 
\quad\text{with}\quad r=-\frac{\Gamma}{K+\Gamma'}
\end{equation}
derived in Ref.~\onlinecite{Cha15}. As we find, the quantum fluctuations
included in the exact groundstate push the ordered moments much closer to the
honeycomb plane. The difference is substantial and needs to be considered when
trying to quantify the model parameters based on the experimental data.

So far, we have considered $\Gamma'=0$ only, while a small negative $\Gamma'$
is expected to be generated by a trigonal compression \cite{Bha12,Rau14b,Cha15}. 
Based on Eq.~\eqref{eq:Eclxypl}, $\Gamma'$ is expected to effectively shift 
the value of $\Gamma$ in the first approximation. Indeed, as shown in
Fig.~\ref{fig:Gdep}(d), the rough three-phase picture as in
Fig.~\ref{fig:Gdep}(a) is preserved and the negative $\Gamma'$ shifts the
$\alpha(\Gamma)$ curve in the negative direction. This enables $\alpha$ to
reach higher values, even above the \mbox{$xy$-plane} angle $54.7^\circ$.

Finally, in Fig.~\ref{fig:Gdep}(e) we briefly analyze the AF $K$ situation
with the moment near the \mbox{$z$-axis}. In contrast to the FM $K$ case, small
$\Gamma$ has a relatively little effect here, because the \mbox{$z$-axis} is
classically selected by the dominant $K>0$ itself. Quantum fluctuations are
found to generate an even stronger pinning to the \mbox{$z$-axis}, compared to the
classical solution of Ref.~\onlinecite{Cha15}. Only a very large $\Gamma$ 
coupling is able to take the spin away from the \mbox{$z$-axis}.



\section{Comparison to experiment}
\label{sec:exper}

\subsection{Extracting pseudospin direction from resonant x-ray and neutron scattering data}

Having quantified the pseudospin easy axis direction as a function of the 
Hamiltonian parameters, we consider now how this ``pseudomoment'' direction 
is related to that of real magnetic moments measured by neutron and
\mbox{x-ray} scattering experiments. To this end, we first define the 
pseudospin one-half wavefunctions including crystal field of trigonal 
symmetry. The latter splits the $t_{2g}$ manifold into an orbital singlet
$a_{1g}=\frac1{\sqrt3}(xy+yz+zx)$, and the 
$e_g'$ doublet $\bigl\{ \frac1{\sqrt6}(yz+zx-2xy)
\: ;\; \frac1{\sqrt2}(zx-yz) \bigr\}$. 
Denoting this splitting by $\Delta$ and using the hole-representation, 
we have: 
\begin{equation}
H=\Delta\,\tfrac13 \left[2n(a_{1g})-n(e_g')\right] \;.
\end{equation}
Within a point-charge model, positive (negative) $\Delta$ would correspond to
a compression (elongation) of octahedra along the trigonal \mbox{$c$-axis}.
The actual value of $\Delta$ in real material is decided by various factors,
but this issue is not relevant in the present context. 

In terms of the effective angular momentum $l=1$ of the $t_{2g}$ shell, 
$a_{1g}$ state corresponds to the $l_c=0$ state, while the $e_g'$ doublet hosts
the $l_c=\pm 1$ states, using the quantization axis $c$ suggested by the
trigonal crystal field. Explicitly,
\begin{align}
|0\rangle &= \frac1{\sqrt3} (|yz\rangle + |zx\rangle +|xy\rangle) \;, \\
|\pm 1\rangle &= \pm\frac1{\sqrt3}(\mathrm{e}^{\pm {2\pi i}/3} |yz\rangle 
+\mathrm{e}^{\mp {2\pi i}/3} |zx\rangle +|xy\rangle) \;.
\end{align}

Via these $l_c$ states, pseudospin-$\frac12$ wavefunctions are defined as:
\begin{align}
|+\tfrac12\rangle &=+\sin\vartheta\, |0,\uparrow\rangle 
-\cos\vartheta\, |+1,\downarrow\rangle \;, \label{eq:pseudoA}\\
|-\tfrac12\rangle &=-\sin\vartheta\, |0,\downarrow\rangle 
+\cos\vartheta\, |-1,\uparrow\rangle \;, \label{eq:pseudoB}
\end{align}
where $\uparrow$ and $\downarrow$ refer to the projections of the hole spin on
the trigonal \mbox{$c$-axis}. The spin-orbit ``mixing'' angle 
$0\le \vartheta \le \pi/2$ is given by
$\tan 2\vartheta = {2\sqrt2}/{(1+\delta)}$, where $\delta=2\Delta/\lambda$. 

Using the wavefunctions \eqref{eq:pseudoA} and \eqref{eq:pseudoB}, we may 
express the spin $\vc s$ and orbital $\vc l$ moments of a hole via the 
pseudospin $\vc S$. In a cubic limit, i.e. $\Delta=0$, one has 
$\vc s=-\frac13 \vc S$, $\vc l=\frac43 \vc S$, and total magnetic moment 
$\vc M=(2\vc s -\vc l)=-2 \vc S$ (note a negative $g$-factor $g=-2$). These
relations imply that the pseudospin easy axis direction is identical to that 
of spin, orbital, and magnetic moments when trigonal field is zero. However, 
this is no longer valid at finite $\Delta$. For instance, strong compression 
($\vartheta =0$) would completely suppress the \mbox{$ab$-plane}
components of magnetic moments, so the pseudospin and magnetic moment 
will not be parallel anymore (unless pseudospin is ordered along the 
\mbox{$c$-axis}). 

The x-rays and neutrons couple initially to the spin and orbital moments, and
the scattering operator has to be projected onto the pseudospin basis. We
first consider an effective RXS operator. For pseudospin one-half in a
trigonal field, it has to have a form 
$\hat{R}\propto i f_{ab}(P_aS_a+P_bS_b) +i f_cP_cS_c$, where 
$\vc P = \vc\varepsilon\times\vc\varepsilon'$ and $\vc\varepsilon$ 
($\vc\varepsilon'$) is the polarization of the incoming (outgoing) photon. 
This can be written as $\hat{R}\propto i \vc P \cdot \vc N$, introducing 
a vector $\vc N=(f_aS_a, f_bS_b, f_cS_c)$ with $f_a=f_b\equiv f_{ab}$. The 
RXS data determines a direction of this auxiliary vector $\vc N$; in
Na$_2$IrO$_3$, it was found to make an angle $\alpha_N\approx 44.3^\circ$ to
the \mbox{$ab$-plane} \cite{Chu15}. However, this is not yet the pseudospin 
direction, since $f_{ab} \neq f_c$ and hence $\alpha_S \neq \alpha_N$, unless 
the trigonal field is exactly zero (unlikely in real materials). To access 
the pseudospin angle $\alpha_S$ and quantify the model parameters, one 
has to know the ``RXS-factors'' $f_{ab}$ and $f_c$.

\begin{figure}[t!]
\begin{center}
\includegraphics[scale=1.00]{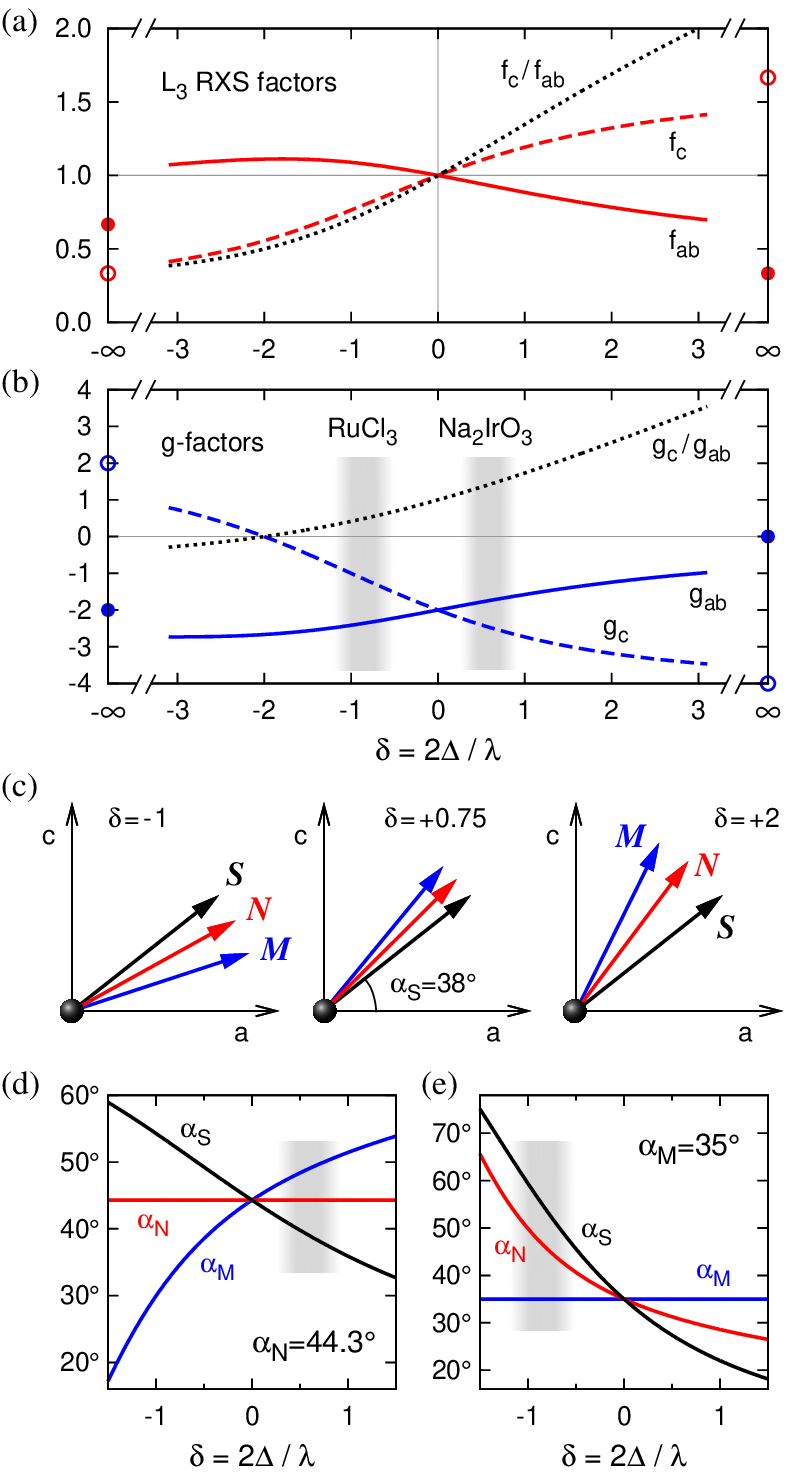}
\caption{(Color online)
(a)~Factors $f$ entering the relation between the pseudospin $\vc S$ and $L_3$
RXS vector $\vc N$ presented as functions of the trigonal field.
(b)~$g$-factors as functions of the trigonal field. Intervals of $\delta$
consistent with the \mbox{$g$-factors} suggested by the experimental data on
\mbox{RuCl$_3$} \cite{Kub15,Joh15} and \mbox{Na$_2$IrO$_3$}
\cite{Sin10,Gre13} are indicated by shading.
(c)~Directions of the $\vc S$, $\vc N$, and $\vc M$ vectors for sample values
of the trigonal field parameter $\delta$ and a fixed pseudospin angle
$\alpha_S=38^\circ$. The case with the negative $\delta=-1$ 
could be relevant for RuCl$_3$, while positive $\delta=+0.75$ with the 
reverse order of the vectors $\vc M$, $\vc N$, and $\vc S$ for Na$_2$IrO$_3$.
(d),(e)~Angles $\alpha_S$, $\alpha_N$, and $\alpha_M$ of the vectors $\vc S$,
$\vc N$, and $\vc M$ to the honeycomb plane as functions of $\delta$ keeping
fixed $\alpha_N=44.3^\circ$ (d) or $\alpha_M=35^\circ$ (e). The shaded
$\delta$-intervals are the same as in panel (b).
}
\label{fig:SNM}
\end{center}
\end{figure}

We have derived the $f$-factors (see Appendix \ref{app:RXS} for details). 
For the $L_3$ edge, they read as: 
\begin{align}
f_{ab} &= \frac12 + \frac5{6\sqrt2}\, s_{2\vartheta} - \frac16\,
c_{2\vartheta} \;, \\
f_c &= 1 + \frac23\, c_{2\vartheta} - \frac1{3\sqrt2}\, s_{2\vartheta} \;.
\end{align}
Here, $s_{2\vartheta}=2\sqrt2/r$, $c_{2\vartheta}=(1+\delta)/r$, and 
$r=\sqrt{8+(1+\delta)^2}$. Fig.~\ref{fig:SNM}(a) shows the $f$-factors  
as a function of trigonal field parameter $\delta$. In cubic limit, one 
has $f_{ab}=f_c$ hence $\vc N$ is parallel to $\vc S$, as expected.

For completeness, we show also the $f$-factors for the $L_2$ edge:
\begin{align}
f_{ab}=2f_c =-\frac32 +\frac12\, c_{2\vartheta} +\sqrt{2}\, s_{2\vartheta} \;, 
\end{align}
which vanish at $\delta=0$ limit, as a consequence of the spin-orbit 
entangled nature of pseudospins \cite{Kim09}. 

In neutron diffraction experiments, the magnetic moment 
$\vc M=(g_a S_a,g_b S_b,g_c S_c)$ is probed. For the pseudospins as defined 
above, the $g$-factors are (neglecting covalency effects \cite{Abr70}): 
\begin{align}
g_{ab}&= -(1 + \sqrt2\, s_{2\vartheta} - c_{2\vartheta}) \;, \\
g_c &= -(1 + 3\, c_{2\vartheta}) \;.
\end{align}
The $g$-factor anisotropy can quantify the strength of the trigonal field, as
illustrated in Fig.~\ref{fig:SNM}(b). Again, magnetic moment direction is 
in general different from that of pseudospin, and to access the latter one
needs to know the $g$-factors. 

These considerations imply that the orientations of the \mbox{(x-ray)} $\vc N$
vector and magnetic moment $\vc M$ differ from each other, and also from 
that of pseudospin $\vc S$ which enters the model Hamiltonian. As we show in 
Fig.~\ref{fig:SNM}(c), their relative angles come in the order 
$\alpha_M>\alpha_N>\alpha_S$ for positive $\Delta$, and in reversed order 
$\alpha_S>\alpha_N>\alpha_M$ for negative $\Delta$. Ideally, having 
measured both $\vc N$ and $\vc M$ directions in the same compound, one could
extract the crystal field parameter $\delta$ using the above equations, and
uniquely fix the pseudospin easy axis angle $\alpha_S$. In principle,
the $g$-factor anisotropy provides the same information on $\delta$, but 
obtaining $g$-factors in magnetically concentrated systems is somewhat
nontrivial task. Alternatively, one could extract the value and sign of
$\Delta$ directly from the splitting and anisotropy of high-energy $J=3/2$
quartet in single crystals (see Appendix \ref{app:trans} for details).


\subsection{Implications for Na$_\mathbf{2}$IrO$_\mathbf{3}$ and RuCl$_\mathbf{3}$}

Armed with the above relations between different moments, and using the
results of Sec.~\ref{sec:zz}, let us now analyze the available experimental
data on Na$_2$IrO$_3$ and RuCl$_3$.

Starting with the case of Na$_2$IrO$_3$, we utilize the value $\alpha_N\approx
44.3^\circ$ determined recently by RXS \cite{Chu15}. Keeping this experimental
constraint, in Fig.~\ref{fig:SNM}(d) we plot the remaining angles $\alpha_M$
and $\alpha_S$ as functions of the relative strength of the trigonal crystal
field $\delta$. In Ref.~\onlinecite{Cha15}, the value 
$\Delta/\lambda\approx 3/8$ was deduced based on the splitting
$\Delta_{BC}\approx 0.1\:\mathrm{eV}$ of $J=3/2$ quartet \cite{Gre13}. As
seen in Fig.~\ref{fig:SNM}(b), the corresponding $\delta\approx 0.75$ is also
roughly consistent with the anisotropy of the \mbox{$g$-factors},
$g_c/g_{ab}\approx 1.4$, obtained by fitting the temperature-dependent
magnetic susceptibilities $\chi_{c}>\chi_{ab}$ \cite{Sin10}. The data in
Fig.~\ref{fig:SNM}(d) then suggests that the magnetic moment takes an angle of
about $\alpha_M\approx 50^\circ$ to the honeycomb plane, while the pseudospin
angle $\alpha_S$ is roughly $38^\circ-40^\circ$. Such a deviation of the
pseudospin from the \mbox{$xy$-plane} ($\alpha\approx 54.7^\circ$) implies
a sizable
$\Gamma$ value. Based on Fig.~\ref{fig:Gdep}(c) we may naively expect the
$\Gamma/|K|$ ratio in the range $0.3-0.5$. We emphasize, however, that this
conclusion relies on the above estimate of the trigonal field, that should be
verified by measuring the ``magnetic'' angle $\alpha_M$ directly by neutron
scattering.

Compared to Na$_2$IrO$_3$, RuCl$_3$ shows an opposite magnetic anisotropy
behavior with $\chi_{c}\ll\chi_{ab}$ \cite{Kub15}. The magnetic structure has
been recently investigated by neutron scattering \cite{Cao16}, with the result
$\alpha_M\approx 35^\circ$ and $\varphi$ being equal to either $0^\circ$ or
$180^\circ$. Similarly to Fig.~\ref{fig:SNM}(d), in Fig.~\ref{fig:SNM}(e) we
keep the measured angle, now $\alpha_M$, fixed at its experimental value, and
plot $\alpha_S$ and $\alpha_N$ for varying $\delta=2\Delta/\lambda$. This
parameter could be obtained from the anisotropy of $J=3/2$ transitions in
single crystals (see Appendix \ref{app:trans}). We are not aware of such a
direct measurement in RuCl$_3$, so the trigonal field is best
assessed by considering the anisotropy of the \mbox{$g$-factors}.
Refs.~\onlinecite{Kub15,Joh15} reported in-plane and out-of-plane
magnetization curves measured for high fields up to $60\:\mathrm{T}$. Even
though the saturation was not reached, the data indicate the value
$g_c/g_{ab}\approx 0.4-0.5$. A similar ratio was also found by Yadav 
{\it et al.} \cite{Yad16} using quantum chemistry methods and by fitting the
high-field data of Ref.~\onlinecite{Joh15}. The corresponding $\delta$ puts
the pseudospin angle $\alpha_S$ at relatively high values of about $\alpha_S
\gtrsim 50^\circ$ [see Fig.~\ref{fig:SNM}(e)]. Adopting this estimate, we will
try to identify a consistent parameter window.

Unfortunately, the present neutron experiment \cite{Cao16} could not directly
resolve the orientation of the moments with respect to the \mbox{$a$-axis},
i.e. whether $\varphi=0^\circ$ or $\varphi=180^\circ$. The absence of this
most conclusive evidence for the sign of the Kitaev interaction requires us to
consider both possibilities.

We assume first FM $K<0$ as obtained in two recent ab-initio calculations of
the exchange interactions in RuCl$_3$ \cite{Win16,Yad16}.
Fig.~\ref{fig:Gdep}(c) gives a hint that the estimated $\alpha_S \gtrsim
50^\circ$ can be reached for small $\Gamma$ only. As seen in
Fig.~\ref{fig:Gdep}(d), by including small negative $\Gamma'$ that shifts the
crossover towards negative $\Gamma$, the pseudospin direction may rotate even
far above the \mbox{$xy$-plane}. Interestingly, the corresponding parameter
regime $J\sim -\Gamma \sim -\Gamma' \sim 0.2|K|$ matches well the prediction
by quantum chemistry calculations \cite{Yad16}. 

Now we analyze the AF $K>0$ case, proposed for RuCl$_3$ in
Refs.~\onlinecite{Kim15,Cao16,Ban16}. In this case, the zigzag order is
obtained on the level of the two-parameter Kitaev-Heisenberg model
\cite{Cha13} alone, and this simplicity makes the AF $K$ scenario particularly
attractive. In the zigzag phase of the two-parameter model, the pseudospins
point along the cubic \mbox{$z$-axis} leading to $\alpha_S\approx 35^\circ$.
This can be reconciled with the experimental value $\alpha_M\approx 35^\circ$
only in a nearly cubic situation with a small trigonal distortion. Considering
however the large anisotropy of the $g$-factors discussed above and the
resulting $\alpha_S\gtrsim 50^\circ$, it seems that the AF Kitaev interaction
needs to be supplemented by other anisotropic interactions lifting the
pseudospin considerably up. This scenario is addressed in
Fig.~\ref{fig:Gdep}(e). We have found, that $\Gamma'$ does not influence
$\alpha_S$ much so that we focus on the $\Gamma$-dependence. Since the AF $K$
zigzag phase becomes fragile if the other anisotropy terms are included, the
model has to be additionally extended by $J_3$. Based on the data of
Fig.~\ref{fig:Gdep}(e), we may conclude that large negative $\Gamma$
comparable to $K$ is needed to obtain $\alpha_S\gtrsim 50^\circ$. It should be
carefully checked if such a substantially extended model is still consistent
with other experimental data, in particular with the spin excitation spectrum
with small only gaps \cite{Ban16}.

We would like to stress again, that our analysis of RuCl$_3$ for both $K<0$
and $K>0$ heavily relied on the relative trigonal field strength
$\Delta/\lambda$ inferred solely from the magnetization anisotropy in high
magnetic fields. It is thus highly desirable to measure the complementary
angle $\alpha_N$ by RXS and quantify $\Delta/\lambda$ more precisely, as
suggested in the previous subsection. As discussed in 
Appendix~\ref{app:trans}, measuring the anisotropy of $J=3/2$ states by
inelastic neutron scattering in single crystals would be also very helpful.

To summarize this section, in Na$_2$IrO$_3$, the measured moment direction
\cite{Chu15} with $\varphi=0^\circ$ well fixes the FM sign of the Kitaev
interaction, and our analysis of its angle from the \mbox{$ab$-plane} suggests
that $\Gamma \sim 0.3-0.5|K|$ coupling is present. Concerning RuCl$_3$, the
current ambiguity in the angle $\varphi$ ($0^\circ$ or $180^\circ$) leaves
open the issue of the sign of $K$. There is also an uncertainty in the
trigonal field value $\Delta$; based so far on the $g$-factor anisotropy, we
found that FM $K<0$ with relatively small $\Gamma$, $\Gamma'$ values would be
consistent with the data, while AF $K>0$ situation requires large $\Gamma<0$
couplings comparable to~$K$. 



\section{Conclusions}

We have investigated the ordered moment direction in the zigzag phases of the
extended Kitaev-Heisenberg model for honeycomb lattice magnets. Our method
analyzes the exact cluster groundstates using a particular set of spin
coherent states and as such fully accounts for the quantum fluctuations.  
The interplay among the various anisotropic interactions leads to a complex
behavior of the ordered moment direction as a function of the model
parameters. We have found substantial corrections to the results of a
classical analysis that are important when quantifying the exchange
interactions based on the experimental data.

We have pointed out that, away from the ideal cubic situation, the notion of
the ``ordered moment direction'' has to be precisely specified. Assuming a
trigonal field relevant to the layered honeycomb systems, we have derived
relations among the directions of \textit{(i)}~the pseudospins entering the
model Hamiltonian, \textit{(ii)}~the magnetic moments measured by neutron
diffraction, and \textit{(iii)}~the moment direction as probed by resonant
magnetic \mbox{x-ray} scattering. These relations and a combination of neutron
and \mbox{x-ray} data should enable a reliable quantification of the trigonal
field as well as the pseudospin direction in future experiments.

Using the above results, we have analyzed the currently available experimental
data on Na$_2$IrO$_3$ and RuCl$_3$ and identified plausible parameter regimes
in these compounds.


\acknowledgments

We would like to thank G.~Jackeli, B.J.~Kim, S.E.~Nagler, and J.~Rusna\v{c}ko for helpful
discussions. 
JC acknowledges support by Czech Science Foundation (GA\v{C}R) under project
no. GJ15-14523Y and M\v{S}MT \v{C}R under NPU II project CEITEC 2020 (LQ1601).



\appendix

\section{Comparison of numerical methods}
\label{app:numer}

As mentioned in the main text, the standard method to obtain the ordered
moment direction using the ED groundstate is to evaluate the spin-spin 
correlation matrix 
$\langle S^\alpha_{-\vc Q} S^\beta_{\vc Q}\rangle$ ($\alpha,\beta=x,y,z$)
at the ordering vector $\vc Q$ and to find its eigenvector corresponding to
the largest eigenvalue.
However, there are two main problems associated with this simple method, both
emerging since the cluster groundstate is a linear superposition of degenerate
orderings where the individual orderings have equal weights:

\textit{(i)} If there are several equivalent easy axis directions associated 
with the selected ordering vector $\vc Q$, they will be characterized by the 
same eigenvalue. This leads to a degenerate eigenspace and prevents us to 
resolve such directions. The most severe cases are those with a dominant
Heisenberg interaction presented in Fig.~\ref{fig:prob}(a,b). Here we have 
three degenerate easy axes $x$, $y$, $z$ which makes the correlation matrix
proportional to a unit matrix and thus isotropic. In the FM $K<0$ zigzag
situation shown in Fig.~\ref{fig:prob}(d) and the entire middle phase in
Fig.~\ref{fig:Gdep}(a), two degenerate moment directions for a particular
zigzag pattern (selected by $\vc Q$) are possible and the correlation matrix
therefore just uncovers the softness of the $xy$-plane. Only after these two
directions merge a for large enough $|\Gamma|$, the moment direction can be
identified.

\textit{(ii)} The zigzag pattern to be probed is selected by choosing the 
ordering vector $\boldsymbol Q$. In contrast to an infinite lattice, at a 
finite cluster this separation of the three zigzag directions is not perfect.
The range of spin correlations is limited by the size of the cluster and the
corresponding momentum space peaks become broad. The correlation matrix at
given $\boldsymbol Q$ is thus ``polluted'' by small contributions of the two
other zigzags in the groundstate, that are associated with the remaining
ordering vectors.

Our method introduced in Sec.~\ref{sec:method} does not suffer from the above
problems and is able to handle all the situations encountered. This is due to
the full resolution of the various degenerate orderings present in the cluster
groundstate by using a prescribed ordering pattern and by a construction of a
full directional map.

\begin{figure}
\begin{center}
\includegraphics[scale=1.00]{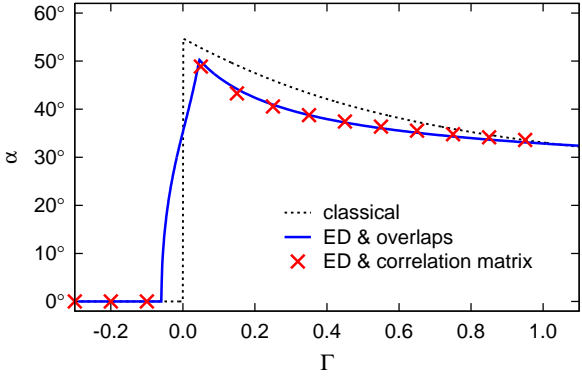}
\caption{(Color online)
Comparison of the angle $\alpha$ of the pseudospin direction to the $ab$-plane
obtained using various methods. The parameters $K\!=\!-1$ and $J\!=\!J_3\!=\!0.2$ 
were used. The blue curve is identical to the one shown in Fig.~\ref{fig:Gdep}(a-d).
}
\label{fig:corrmtx}
\end{center}
\end{figure}

If applicable, the standard method gives results very similar to our method.
We demonstrate this in Fig.~\ref{fig:corrmtx} that compares the two methods
for the parameters $K=-1$, $J=J_3=0.2$ and varying $\Gamma$ used in
Fig.~\ref{fig:Gdep}. The slight deviations observed for $\Gamma>0$ can be
interpreted as a manifestation of the second problem discussed above.


\section{Derivation of the L-edge RXS operator}
\label{app:RXS}

Resonant x-ray scattering is conceptually similar to the Raman light
scattering, in a sense that both processes involve the intermediate states
created and subsequently eliminated by incoming and outgoing photons. However,
the nature of the intermediate states in these two cases is radically
different: while the Raman light scattering involves intersite $d-d$
transitions, the \mbox{x-rays} create the high-energy on-site $p-d$
transitions. As a result, the Raman light scattering probes intersite
(two-magnon) spin flips, while the presence of strong spin-orbit coupled
$2p$-core hole in the RXS intermediate states makes a single-ion spin flips a
dominant magnetic scattering channel (see the recent review \cite{Ame11a} and
references therein for details). 

A complex time-dynamics of the intermediate states makes the \mbox{x-ray}
scattering process hard to analyze microscopically. However, as far as one is
concerned with the low-energy excitations in Mott insulators, the problem of
the intermediate states can be disentangled and cast in the form of frequency
independent phenomenological constants \cite{Ame10,Hav10,Sav15}. This results
is an effective RXS operator formulated in terms of low-energy (orbital, spin,
\ldots) degrees of freedom alone. The form of this operator is dictated by
symmetry. In essence, this approach is similar to that of Fleury and Loudon
\cite{Fle68} widely used in the theories of Raman light scattering in quantum
magnets. 

While the RXS operator used in the main text follows from an underlying
trigonal symmetry, the ratio between $f_{ab}$ and $f_c$ constants requires
specific calculations. This can be easily done, with some routine
modifications of the previous work for the case of tetragonal symmetry
\cite{Ame11b,Kim12}, as outlined below.      
   
In cubic axes $x,y,z$ (see Fig.~\ref{fig:schem}), a dipolar $2p$ to $5d$ 
transition operator reads as: 
\begin{equation}
D=\varepsilon_x T_x + \varepsilon_y T_y+ \varepsilon_z T_z \;,
\end{equation}
where $\varepsilon_{x,y,z}$ are the polarization factors, and 
$T_x=d^\dagger_{zx}p_z+d^\dagger_{xy}p_y$,
$T_y=d^\dagger_{xy}p_x+d^\dagger_{yz}p_z$, 
$T_z=d^\dagger_{yz}p_y+d^\dagger_{zx}p_x$.
Here and below, it is implied that $d$ and $p$ operators carry also
the spin quantum numbers ($\uparrow$, $\downarrow$) over which summation 
is taken.

In the quantization axes $a,b,c$, suggested by the trigonal crystal field, 
this operator takes the following form:
\begin{equation}
D=\frac1{\sqrt{6}}(\varepsilon_a T_a +\varepsilon_b T_b+\varepsilon_c T_c) \;, 
\end{equation}
where 
\begin{align}
T_a&=(d^\dagger_0\!+\!2d^\dagger_{-1})p_1
\!+\!(d^\dagger_1\!-\!d^\dagger_{-1})p_0
\!+\!(2d^\dagger_1\!-\!d^\dagger_0)p_{-1} \;, \notag \\
iT_b&=(-d^\dagger_0\!+\!2d^\dagger_{-1})p_1
\!+\!(d^\dagger_1\!+\!d^\dagger_{-1})p_0
\!-\!(2d^\dagger_1\!+\!d^\dagger_0)p_{-1} , \notag \\
T_c&=\sqrt{2}\,(2d^\dagger_0 p_0\!-\!d^\dagger_1 p_1\!-\!d^\dagger_{-1}
p_{-1}) \;.\label{eq:T}
\end{align}
Here, the indices $0$ and $\pm 1$ stand for the $l_c$ orbital quantum numbers 
of $d$ and $p$ electrons.

Within the above Fleury-Loudon-like approach to the \mbox{x-ray} scattering
problem, effective RXS operator is given by
$D^\dagger(\varepsilon')D(\varepsilon)$, and its part responsible for the
magnetic scattering reads as $\hat{R}\propto
i(\vc\varepsilon\times\vc\varepsilon')\cdot (\vc T^\dagger\times\vc T)$. 

Next, the core-hole operators $p$ in \eqref{eq:T} are expressed in
terms of spin-orbit split $j=1/2$ and $j=3/2$ eigenstates of the $2p$ level,
resulting in two sets of $\vc T$ operators active in $L_2$ and $L_3$ edges,
correspondingly. After ``integrating out'' these $2p_{\frac12}$ and
$2p_{\frac32}$ operators, the product $(\vc T^\dagger\times\vc T)$ becomes a
simple quadratic form of $d$ operators. Finally, projecting this
form onto a pseudospin doublet (given by Eqs. \ref{eq:pseudoA} and
\ref{eq:pseudoB} of the main text), we arrive at the RXS operator
$\hat{R}\propto i f_{ab}(P_aS_a+P_bS_b) +i f_cP_cS_c$, with the $f$-factors
shown in the main text. Via the pseudospin wavefunctions, the RXS $f$-factors
are sensitive to a trigonal field strength. 


\section{Determination of the trigonal field from $\vc{J}\mathbf{=3/2}$ magnetic excitation spectra}
\label{app:trans}

\begin{figure}
\begin{center}
\includegraphics[scale=1.00]{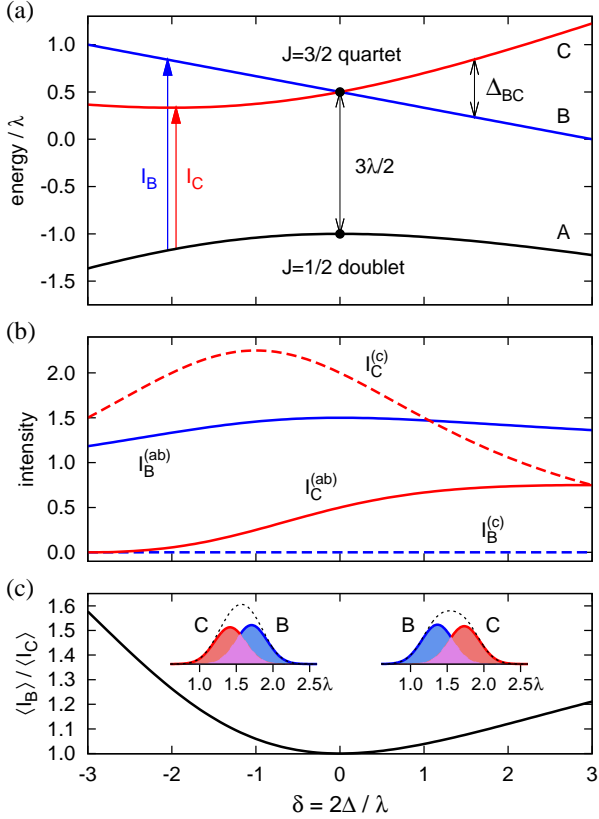}
\caption{(Color online)
(a)~Level structure of a $d^5(t_{2g})$ ion upon trigonal field splitting
characterized by $\delta=2\Delta/\lambda$ (hole picture).
(b)~Intensities of the magnetic transitions $A\!\rightarrow\!B$ 
and $A\!\rightarrow\!C$ for the \mbox{$ab$-plane} and \mbox{$c$-axis} 
components of the dynamical spin structure factor
as given by Eqs. \ref{eq:IB} and \ref{eq:IC}.
(c)~Ratio of the powder-averaged intensities. The insets show the broadened 
(HWHM=$\frac14\lambda$) peak structure assuming $\delta=-1$ (left) and 
$\delta=+1$ (right), respectively.
}
\label{fig:ABC}
\end{center}
\end{figure}

Under spin-orbit coupling $\lambda$ and trigonal crystal field $\Delta$,
$t_{2g}$-hole states split into three levels $A$, $B$, and $C$, see
Fig.~\ref{fig:ABC}(a). The $A$ level hosts a Kramers pseudospin one-half
(corresponding to $J=1/2$ in the cubic limit), with the wavefunctions 
\begin{align}
|A_+\rangle &=+\sin\vartheta\, |0,\uparrow\rangle 
-\cos\vartheta\, |+1,\downarrow\rangle \;,\\
|A_-\rangle &=-\sin\vartheta\, |0,\downarrow\rangle 
+\cos\vartheta\, |-1,\uparrow\rangle \;, 
\end{align}
as were given by Eqs.~\ref{eq:pseudoA} and \ref{eq:pseudoB} of the main text. 
The upper Kramers doublets $B$ and $C$ are derived from spin-orbit $J=3/2$ 
quartet. The former correspond to pure $J_c=\pm 3/2$ states of $J=3/2$ moment:
\begin{align}
|B_+\rangle &= |+1,\uparrow\rangle \;,\\
|B_-\rangle &= |-1,\downarrow\rangle \;, 
\end{align}
while the $C$ level wavefunctions are given by 
\begin{align}
|C_+\rangle &=\cos\vartheta\, |0,\uparrow\rangle 
+\sin\vartheta\, |+1,\downarrow\rangle \;,\\
|C_-\rangle &=\cos\vartheta\, |0,\downarrow\rangle 
+\sin\vartheta\, |-1,\uparrow\rangle \;, 
\end{align}
corresponding to $J_c=\pm1/2$ states of $J=3/2$ quartet in the cubic limit,
and containing some admixture of the original $J=1/2$ doublet at finite
$\Delta$. The energies of these states are: 
$E_{A,C} / \lambda = \frac14[\mp\sqrt{8+(1+\delta)^2}-1] +\frac1{12} \delta$
and
$E_B / \lambda = \frac12 -\frac1{6} \delta$.

Transitions from the ground state $A$ level to $B$ and $C$ states are
magnetically active; their spectral weights in the dynamical spin structure
factor are determined by matrix elements of the magnetic moment 
$\vc M=2\vc s-\vc l$: 
\begin{align}
\mp \langle B_\pm | M_a | A_\pm \rangle
&\!=\! \tfrac1i \langle B_\pm | M_b | A_\pm \rangle 
\!=\! \cos\vartheta +\!\tfrac1{\sqrt2} \sin\vartheta , \\
\pm \langle C_\mp | M_a | A_\pm \rangle
&\!=\! \tfrac1i \langle C_\mp | M_b | A_\pm \rangle 
\!=\! \tfrac12 ( s_{2\vartheta}\! +\! \sqrt2 c_{2\vartheta} ) \:.
\end{align}
Out-of-plane moment $M_c$ matrix elements between $A$ and $B$ vanish 
(independent of the spin-orbit mixing angle $\vartheta$), while 
\begin{equation}
\langle C_\pm | M_c | A_\pm \rangle = \tfrac32 s_{2\vartheta} .
\end{equation}
In the magnetic excitation spectra, a transition $A\!\rightarrow\! B$ gives 
a peak at the energy 
\begin{equation}
E_B-E_A= \frac{\lambda}{4}[\sqrt{8+(1+\delta)^2}+3-\delta] \;,
\end{equation}
with the following intensities for different components of the 
dynamical spin structure factor
\begin{equation}\label{eq:IB}
I_B = 
\begin{cases}
\tfrac14 (3+c_{2\vartheta}+2\sqrt2 s_{2\vartheta}) & \text{($ab$-plane)} , \\
\rule{0mm}{5mm}0 & \text{($c$-axis)} .
\end{cases}
\end{equation}
The second transition $A\!\rightarrow\! C$ is peaked at the energy 
\begin{equation}
E_C-E_A = \frac{\lambda}{2}\sqrt{8+(1+\delta)^2}
\end{equation}
and has the intensity 
\begin{equation}\label{eq:IC}
I_C = 
\begin{cases}
\tfrac14 (s_{2\vartheta}+\sqrt2 c_{2\vartheta})^2 & \text{($ab$-plane)} , \\
\rule{0mm}{5mm}\tfrac94 s_{2\vartheta}^2 & \text{($c$-axis)} .
\end{cases}
\end{equation}
The $B$ and $C$ peaks are separated by 
$\Delta_{BC}/\lambda=\frac14[\sqrt{8+(1+\delta)^2}-3+\delta]$; 
at small trigonal splitting $\Delta\ll \lambda$, this can be approximated as
$\Delta_{BC}\approx \frac23\Delta$. At positive (negative) $\Delta$, the $B$
peak position is lower (higher) than that of $C$ peak, see 
Fig.~\ref{fig:ABC}(a).

Fig.~\ref{fig:ABC}(b) shows that the intensities of both transitions are
highly anisotropic with respect to \mbox{$ab$-plane} and \mbox{$c$-axis}
polarizations, with the opposite behavior of $B$ and $C$ contributions. The
out-of-plane response is due to the $C$ transition exclusively, while $B$ peak
dominates the \mbox{$ab$-plane} intensity. This should enable to distinguish
them and determine thereby both the sign and value of trigonal field parameter
$\delta$ from a single-crystal, spin-polarized neutron scattering data.

On the other hand, the powder averaged intensities of $B$ and $C$ peaks are
nearly the same for realistic values of $\delta$, see Fig.~\ref{fig:ABC}(c).
Even at $|\delta|=1$, the two peaks may overlap to give a single broad line,
leaving an ambiguity in the sign of parameter $\delta$.





\begin{thebibliography}{99}
\bibitem{noteZZ}
However, the separation of the individual zigzag chain directions is an
order-from-disorder effect, since a linear combination of different zigzag
patterns is a classical groundstate as well.
\end{thebibliography}
\end{document}